\documentclass[11pt,twoside]{article}

\usepackage{asp2004}
\usepackage{epsf}
\usepackage{psfig}
\usepackage{natbib}
\usepackage{graphicx}

\begin{document}

\title{Dust Formation in Massive WR+O Binaries: Recent Results}

\author{Sergey V. Marchenko}
\affil{Department of Physics and Astronomy, Thompson Complex Central
  Wing, Western Kentucky University, Bowling Green, KY 42101-3576, USA}

\author{Anthony F.J. Moffat}
\affil{D\'epartement de physique, Universit\'e de Montr\'eal, C.P. 6128,
  Succ. Centre-Ville, Montr\'eal, QC, H3C 3J7, Canada}

\begin{abstract}
The massive, luminous Population I Wolf-Rayet stars can be considered as stars
with the highest known {\it sustained} mass loss rates. Around 10\% of WR stars may form
carbon-rich dust in their dense and inhomogeneous winds. Though we are yet to find how
dust is formed in such an extremely hostile environment, we have made substantial
progress over the past decade. Here we discuss the results of  recent  high-resolution
mid-infrared imaging of a sample of the most prodigious WR `dustars'.
This allows one to map rapidly changing dust-forming regions and derive some basic properties
of the freshly formed dust.
\end{abstract}

\section{The Wolf-Rayet `Dustars'}

The hot, massive Population I Wolf-Rayet stars are the evolved descendents of O-type stars.
WR stars live on the verge of exploding as Supernovae (even hypernovae), thus providing
vitally important information about the very final stages of massive star evolution, possibly
leading to Gamma-Ray Bursts. The extremely luminous
($L ^> _\sim 10^5 \, L_\odot$) and hot ($T_{eff} \gg 20000\, K$) hydrostatic cores of WR stars drive
fast dense winds: average mass-loss rates $\dot M \sim 10^{-5} \,M_\odot /yr$  and terminal
velocities $v_\infty \sim 1000-4500 \,km/s$.
There are three evolutionary-successive WR phases: WN, WC (carbon-rich, dust-producing class we
are going to discuss here), and WO. Even though present-epoch dust production output for
all Galactic WC stars is $^< _\sim 1\%$ of the total Galactic
rate \citep{dwe85,coh91}, the dust-generating WC stars are regarded as outstanding
for three main reasons: (i) The absolute rate of dust production is extraordinarily high, reaching
$\dot M \sim 10^{-6} M_\odot yr^{-1}$ \citep{vdh87,wil95} in some stars. (ii) The dust is formed in a hot, extremely hostile environment,
posing a formidable theoretical problem. (iii) In the early (age$\sim$1 Gyr) universe, WR stars could be very
common, but unique sources of dust, along with subsequent dust-generating SN events, since WR
stars evolve much more rapidly than any lower-mass stars, commonly associated with dust production
in the `modern' universe. It is not clear how much (and what kind of) dust can be produced in a SN
explosion \citep{dwe04}. However, it is quite clear that the copious amounts of the carbon-rich dust
produced in the WC winds may survive for at least $\sim 10^2$ years \citep{mar02}, thus effectively
reaching (and enriching) the ISM.

Two basic processes of dust formation prevail among the WC stars:

\noindent (i) `single' channel: constant, sustained formation in single
WC stars, only of the coolest (WC9,10 and some WC8) subtypes.  The IR
emission excess, arising from re-radiation of stellar UV photons by the
hot dust and superposed on an underlying hotter stellar emission
component, is in the form of $\sim$black-body radiation at $T_d \sim
1000 - 1600K$ from a shell with inner critical diameter $\sim
0.5-1.5\times 10^3 R_\star$ ($\sim 0.5\ -1.5\times 10^4
R_\odot$) (see  Zubko 1998 and references therein). Presumably, the winds of hotter single WC stars are too
rarified to form dust at a distance where the UV radiation has dropped
sufficiently to allow dust formation to occur.  In any case, even in
cool WC stars, a smooth wind flow will not form dust; clumping is
required for efficient grain growth \citep{che00}. Apparently, it cannot be excluded that
the `single' channel might ultimately involve a binary companion and thus a strong wind-wind
collision zone to facilitate the dust formation

\noindent (ii) `binary' channel: episodic formation in binary WC + O
systems with eccentric orbits, and almost constant dust formation in WC+O binaries
with $\sim$circular orbits (`pinwheel nebulae': see \citep{tut99}). The key factor here is the compression by
wind-wind collision involving H-rich material from the O-star and the
C-rich WC wind. This allows dust formation to occur, which is
dramatically enhanced at each periastron passage.  Among the some 7
systems in which episodic dust formation has been detected so far
\citep{wil95}, all have (confirmed or suspected) long periods of
several years, with no preference for hot- or cool-type WC stars.
Presumably, the dust is formed relatively far downstream along the shock
interface, where the temperature has fallen sufficiently from the
initially extremely high values of $10^{6-7}K$.  The shock cone wraps
around the weaker-wind O-star, so that IR dust emission should arise in
a preferred direction far beyond the O-star, as seen from the WR star.

Hot ($T\leq 1500$ K) circumstellar WR dust was only recently spatially resolved around some
WC+O binaries \citep{tut99,mon99,mar99c,mon02}.
All the above-cited near-IR observations have targeted the hottest dust only. The apparent
sub-arcsecond sizes of the barely resolved hot dust regions made next to impossible any direct
application of quantitative models. The first mid-IR ($\lambda\lambda 8-18\mu m$) images of the
spatially-resolved dust cloud around WR112 \citep{mar02} provided data on:
(a) the temperature profile in the envelope, proving that, as anticipated, the temperature follows
a thermal equilibrium profile;
(b) the characteristic size and chemical composition of the
dust grains, finding amorphous carbon as a main constituent of dust particles
of a $\sim 0.5 \mu m$ characteristic
size; (c) the absolute rates of  dust formation, up to $\dot M(dust) \sim 6\,10^{-7}\, M_\odot \,yr^{-1}$.
Unusually large, sub-micron, dust partilcles were found in independent surveys \citep{chi01,yud01}.

\section{Colliding-Wind Prototypes WR137 and WR140}

\begin{figure}[!ht]
\centering
\plottwo{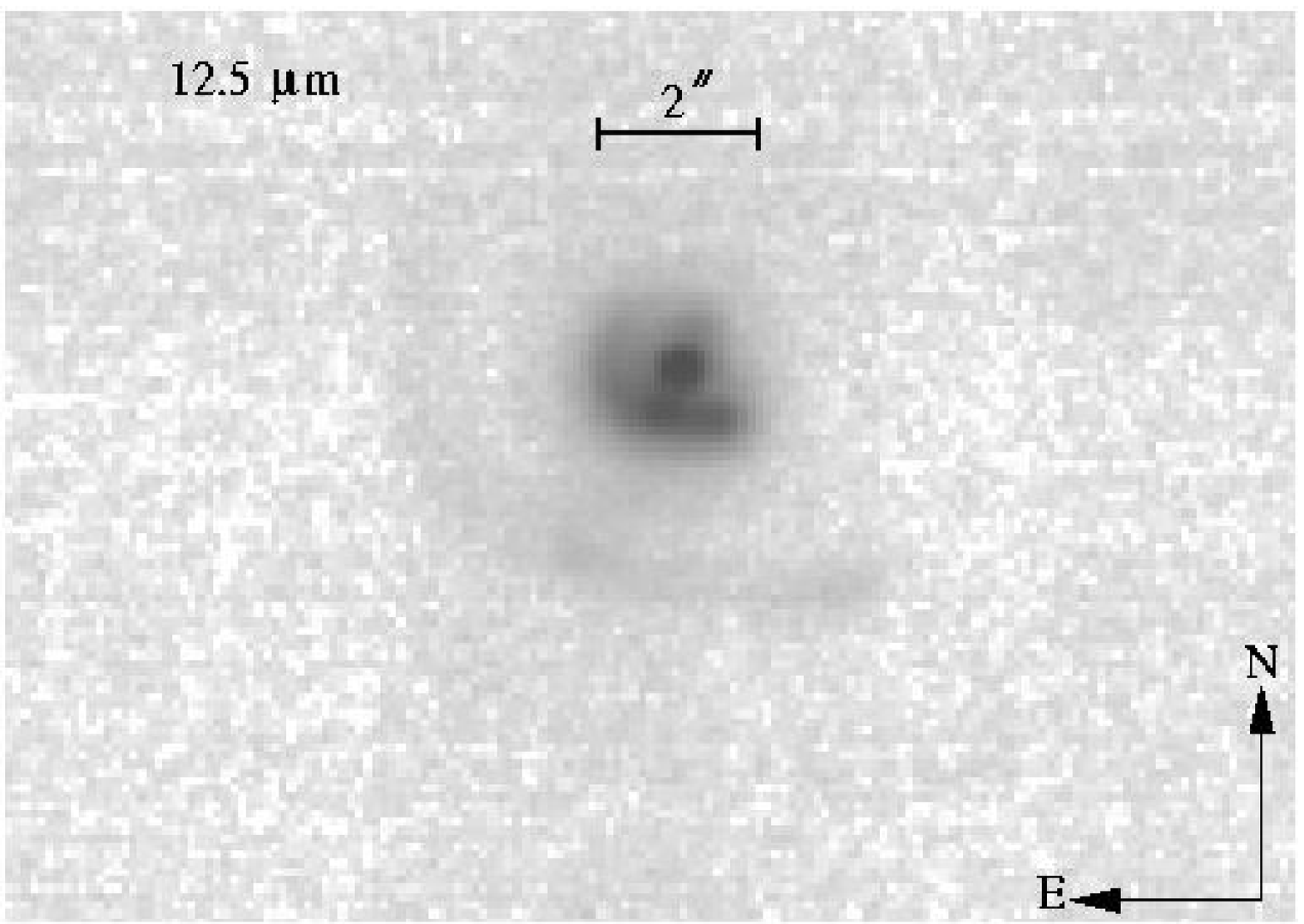}{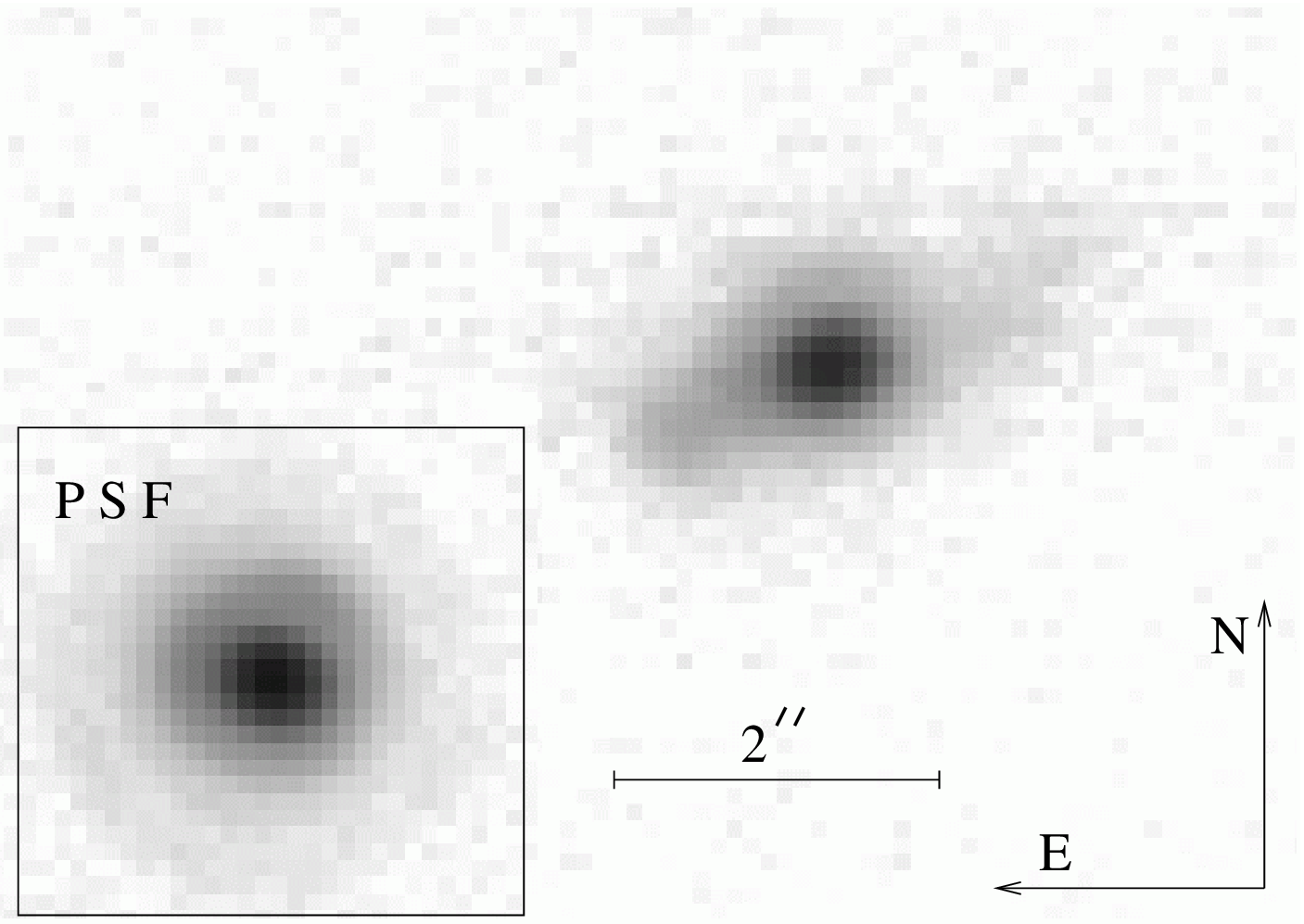}
\caption{The $\lambda 12.5\mu m$ images obtained with Michelle/Gemini-North in November, December 2003.
Left panel: WR 140; right panel: WR 137. }
\end{figure}

The long-period, colliding-wind binaries WR 137 and WR 140 can be used as textbook examples of the
`binary' channel of dust production. WR140: this long-period (P=7.93 y), highly eccentric
(e=0.881: \citet{mar03b}) binary serves as a prototype for studies of wind-wind collision phenomena in massive binaries.
Its repeatable, giant near-IR outbursts \citep{wil90} are related to periastron passages, i.e. the intervals when the O star companion ploughs
through the densest regions of the WR wind. Additional compression of the WR outflow in the wind-wind collision zone
allows the shocked gas to cool rather quickly, thus creating favorable conditions for condensation of dust which eventually
leaves the system. Indeed, after the 2001 periastron passage Monnier et al. (2002) detected slowly [out]moving clumps
of dust. The recent VLBA (S. Dougherty, these proceedings), UKIRT (P. Williams, these proceedings) and IOTA3
\citep{mon04} imaging helped to establish a relation between the 2001 dust ejecta and  the wind-wind collision zone
wrapped around the O star.
Our high-quality $\lambda 12.5\mu m$ images obtained with Michelle/Gemini-North in November, December
2003 show  concentric dust arcs around WR140 which can be unequivocally linked with the 1993 and 2001
dust formation episodes (Fig. 1). Combined with the independently acquired near-IR images, these data may allow
one to confirm the preliminary conclusions about the properties of dust in WR 140, i.e. the presence of
fairly large dust grains with a characteristic size $a \sim 0.1 \mu m$ \citep{mar03b}.

The massive long-period binary WR 137 was recognised as a periodically variable IR source by Williams et al. (1985) and
included in the category of dust-producing WR+O binaries \citep{wil01}. Thirteen years after
the 1984 dust-production maximum,
alarmed by the rising near-IR flux \citep{wil96},
we took H,K images with NICMOS2/HST \citep{mar99c} to find that, roughly one year after the 1996 periastron passage
the system ejected a $\sim$0.25" dust cloud at P.A.$\sim 110^o$ and a bright clump at P.A.$\sim -70^o$, right along the
plane corresponding  to orientation of the equatorially-enhanced wind of the WR component \citep{har00}. Taking the
$\lambda 12.5\mu m$ images in 2003 (Fig. 1) we find that
the dust clouds continue to expand along the directions outlined by the 1997/1998 NICMOS2 images, with the main outburst
going SE. There is no doubt that the equatorially-enhanced mass loss in the WR component plays a leading role in shaping the
outcoming dust cloud.

\section{Mid-IR Survey of Dustars with Gemini}

Finishing our mid-IR survey of the WR `dustars' in 2004, we find a spectacular dust envelope around
WR48a (Fig. 2). This presumably binary system underwent a major dust-formation outburst in 1979 followed
by relatively small secondary outbursts in 1990 and 1994 \citep{wil96}. Can the bright knots in Fig. 2 be related
to the secondary events? If not, then what else can influence the dust production rate at the particular (and repeatable
from orbit to orbit) orbital phases?

Comparing the second-epoch image of WR112 (Fig. 3) taken on July 2004 to the image from May 2001 \citep{mar02},
we find clear signs of slow expansion of the broken dust spiral. The rate of expansion allows us to estimate the distance
to WR112: $d=2.0^{+1.7} _{-0.8} \, kpc$, assuming $v_\infty =1200 \, km\, s^{-1}$. We re-inforce our previous
conclusion that dust may survive in the hostile environment of the WR wind for a long time, up to $10^2$ years.

Six other `dustars': WR 76, WR 80, WR 95, WR 98a,  WR 106 and WR 121 - were not resolved in the $\lambda 12.3\mu m$
images taken with TReCS/Gemini-South.

\begin{figure}[!ht]
\centering
\plotone{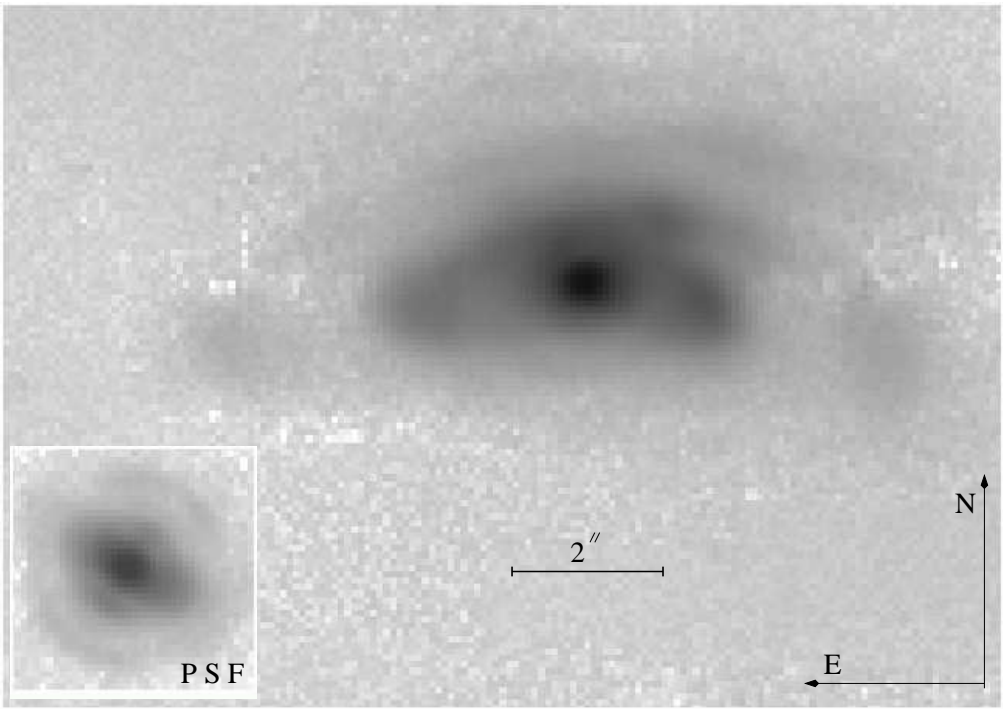}
\caption{The $\lambda 12.3\mu m$ image of WR48a obtained with TReCS/Gemini-South in March 2004.}
\end{figure}

\section{WR Dustars in 2MASS}

We use the excellent sky coverage of 2MASS \footnote{http://www.ipac.caltech.edu/2mass/releases/allsky/} to conduct
a galaxy-wide census of WR stars, adding to the 227 sources listed in  \citep{vdh01} all the recently discovered Galatic Center
objects (see the compilation in \citet{vdh03}). Working with the J-H vs. H-K diagrams of appropriately de-reddened
sources and combining them with the K vs. J-K distributions, we were able to isolate the known population of WR `dustars'
\citep{woo03} and find one new dust-producing source, WR102e. Overall, there is no statistically significant difference
between the known binary and presumably single WR `dustars': $<J-H>_{single}=0.95 \pm 0.09$,
$<H-K>_{single}=0.90 \pm 0.05$, while $<J-H>_{binary}=0.76 \pm 0.17$,
$<H-K>_{binary}=0.83 \pm 0.10$. Slightly bluer color indices of the binaries can be explained by the presence of
a luminous, hot companion.

\begin{figure}[!ht]
\centering
\plotone{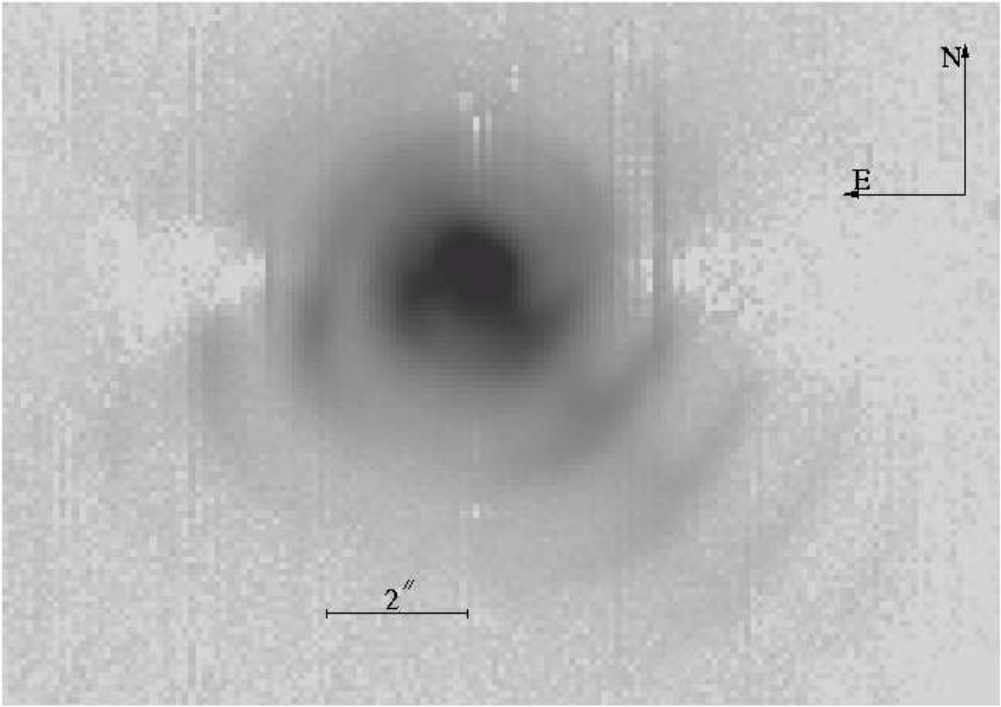}
\caption{The $\lambda 12.3\mu m$ image of WR112 obtained with TReCS/Gemini-South in July 2004.}
\end{figure}



\end{document}